\documentclass[reprint,prd,preprintnumbers,nofootinbib]{revtex4-1}
\usepackage{graphicx,amssymb,amsmath,color}
\usepackage{hyperref}
\usepackage{color}

\def\beq{\begin{equation}}
\def\eeq{\end{equation}}

\def\bea{\begin{eqnarray}}
\def\eea{\end{eqnarray}}
\def\bei{\begin{itemize}}
\def\eei{\end{itemize}}
\def\bmat{\begin{matrix}}
\def\emat{\end{matrix}}
\def\ble{\begin{flushleft}}
\def\ele{\end{flushleft}}
\def\={\,=\,}
\def\+{\,+\,}
\def\-{\,-\,}

\newcommand{\Fig}[1]{Fig.~\ref{#1}}

\newcommand{\Sec}[1]{Sec.~\ref{#1}}

\begin{document}

\title{Constraining Higgsino Kink Tracks from Existing LHC Searches}

\author{Sunghoon Jung}
\email{shjung@slac.stanford.edu}
\affiliation{Korea Institute for Advanced Study, Seoul 02455, Korea}
\affiliation{SLAC National Accelerator Laboratory, Menlo Park, CA 94025, USA}

\author{Hye-Sung Lee}
\email{hlee@ibs.re.kr}
\affiliation{CERN, Theory Division, CH-1211 Geneva 23, Switzerland}
\affiliation{Center for Theoretical Physics of the Universe, Institute for Basic Science (IBS), Daejeon 34051, Korea}

\begin{abstract} 
Considering the supersymmetric model with the long-lived charged Higgsino, we discuss how Higgsino kink signals can show up in the latest LHC Disappearing Track (DT) and stable chargino searches. We derive constraints on the Higgsino kink signal, and characterize it in comparison to the Wino DT and the slepton kink track. We also discuss how to infer Higgsino model kinematics.
\end{abstract}

\preprint{KIAS-P15011, CERN-PH-TH-2015-039}

\maketitle

%\tableofcontents

%%%%%
\section{Introduction}

The ``kink'' signal is often predicted in models beyond the Standard Model (SM). In the gauge-mediated supersymmetric models, for example, the charged slepton next-to-lightest superparticle (NLSP) becomes long-lived due to very weak interactions of the gravitino lightest superparticle (LSP), and it can subsequently decay to a charged lepton within a detector. Since the momentum directions of these charged particles are different in general, tracks will suddenly be deflected (or kinked) at the decay position, and their tracks form a kink track~\cite{Dimopoulos:1996vz,Asai:2011wy,Hamaguchi:2004df}.\footnote{In this paper, we will always call the track with a kink by simply the ``kink'' or the ``kink track'' for our convenience.}

The kink has been searched for at the LEP~\cite{Barate:1999gm,Barate:1998zp}, but no direct constraints on it have been reported from LHC experiments. A recent study by theorists~\cite{Evans:2016zau} claims that the slepton kink can be constrained by Disappearing Track (DT) searches at the LHC~\cite{Aad:2013yna,CMS:2014gxa}. The DT search is aimed to look for a different signal -- the charged Wino decaying inside the LHC tracker to the almost degenerate neutral Wino, producing only an invisibly soft pion~\cite{Thomas:1998wy,Feng:1999fu,Gunion:1999jr,Ibe:2012hu} and leaving no energetic tracks afterwards. But the kink track resembles the DT in the sense that the original track does not extrapolate to the end of detectors. 
 
In this paper, we consider another supersymmetric model where the charged Higgsino can be the kink track. The charged Higgsino is a long-lived NLSP, eventually decaying to a weakly interacting gravitino LSP and a charged lepton, producing a kink signal. We consider this model because this model realized in the MSSM contains light Higgsinos that might be related to the weak-scale hierarchy problem. This model has been studied in only a small number of theoretical papers~\cite{Kribs:2008hq}, and our study derives the latest new constraints on this model. The Higgsino kink also has several features that can be compared with often discussed signals of the Wino DT and the slepton kink. Although the Wino DT produces only invisibly soft pions, the Higgsino kink can accompany energetic leptons and/or quarks (from intermediate $W$ boson decays) moving in different directions from original Higgsino momentum. Also, although the slepton kink decays to only a single lepton (for simplicity, we do not consider the tau decay), the Higgsino kink can decay to multiple quarks as well as to neutrinos. In addition to these differences in decay kinematics, the kinematics of the fermionic Higgsino pair production is different from that of the scalar slepton pair, as will be discussed. We will study how such differences can become apparent and can be utilized in DT searches to search for Higgsino kinks and to characterize them.

The rest of this paper is organized as follow. In \Sec{sec:model}, we introduce the model of the long-lived charged Higgsino NLSP. In \Sec{sec:analysis}, we present our results for how Higgsino kink would show up in the LHC DT observables and discuss how they can characterize similarities and differences among the aforementioned kink and DT models. Then we derive constraints on the Higgsino kink in the later part of \Sec{sec:analysis}, and we conclude in \Sec{sec:conclusion}.

%%%%%
\section{Model -- Charged Higgsino NLSP and Gravitino LSP} \label{sec:model}

We consider the charged Higgsino as NLSP in the minimal supersymmetric standard model (MSSM) extended with the weakly interacting gravitino LSP. The chargino NLSP is long-lived and eventually decays to the gravitino LSP via $\widetilde{H}^\pm \to W^\pm \widetilde{G}$, leaving kink tracks~\cite{Kribs:2008hq}. 

The tree-level mass splitting between the lightest neutral Higgsino and the charged Higgsino is given by \cite{Kribs:2008hq} (for $M_2 > 0$) 
\bea
&&\Delta m_{\rm tree} \= m_{\widetilde{H}^+} - m_{\widetilde{H}_1^0} \label{eq:deltam} \\
&& \quad \approx \left\{ \left( t_W^2 \frac{M_2}{M_1} + 1 \right) \+ \left( t_W^2 \frac{M_2}{M_1} -1 \right) \frac{\mu}{|\mu|} \sin 2\beta \right\} \frac{M_W^2}{2 M_2},  \nonumber
\eea
where $M_{1,2}$ are gaugino mass parameters and $\mu$ is the Higgsino mass parameter.
$t_W \equiv \tan\theta_W$ is the weak mixing parameter, and $\tan\beta \equiv v_u / v_d$ is the ratio of the vacuum expectation values of the two Higgs doublets.
The $\Delta m_{\rm tree}$ is negative when $M_2 > |M_1| (\gg \mu) >0$ and $M_1 <0$ with small $\tan\beta$, so that the charged Higgsino can be the lightest Higgsino components; see Refs.~\cite{Kribs:2008hq,Chun:2016cnm} for more discussions.

\begin{figure}[t] \centering
\includegraphics[width=0.45\textwidth]{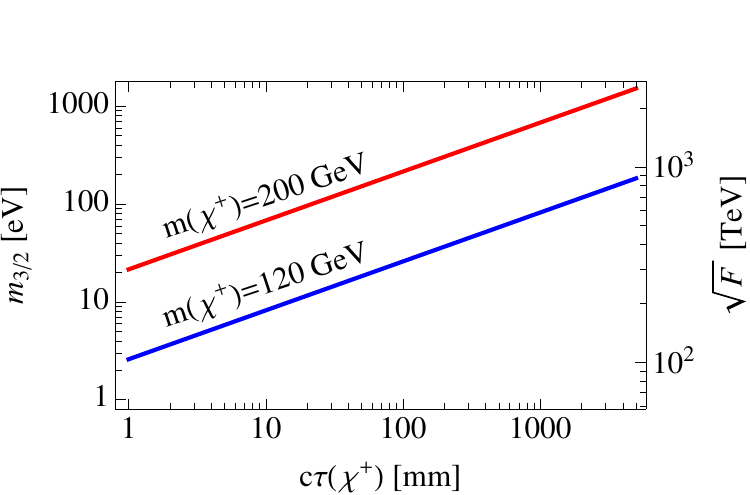}
\caption{Proper decay length of the charged Higgsino NLSP decaying to the gravitino LSP.}
\label{fig:m32}
\end{figure}

The charged Higgsino decay width is~\cite{Ambrosanio:1996jn}
\beq
  \Gamma(\widetilde{H}^+ \to W^+ \widetilde{G}) \= \frac{ m_{\widetilde{H}^+}^5}{32\pi \, F^2 } \, \left( 1- \frac{m_W^2}{m_{\widetilde{H}^+}^2} \right)^4, 
\label{eq:higgsinoctau}\eeq 
where the SUSY breaking scale $F$ is related to the gravitino mass $m_{3/2}$ by $m_{3/2} = \frac{F}{\sqrt{3} M_p}$ with the reduced Planck mass $M_p = 2.4 \times 10^{18}$ GeV. The relevant proper decay length of $c\tau \sim {\cal O}(1-100)$ cm is readily obtained in this model as shown in \Fig{fig:m32}. The needed low scale $F \sim ({\cal O}(100-1000) \, {\rm TeV} )^2$ does not overclose the universe~\cite{Giudice:1998bp}. We take the Higgsino mass and the lifetime $c\tau$ as free parameters.

We assume that slightly heavier neutral Higgsino components promptly decay to the charged Higgsino (the decay to the gravitino LSP is slower); this is realized for $|M_1|,|M_2| \lesssim 2$ TeV and small $\tan \beta$. Thus, the effective production rate of the charged Higgsino pair is a sum of all possible pair production of neutral and charged Higgsinos (and each event eventually contains two long-lived charginos). We have used cross-sections from the LHC Working Group results~\cite{susywg}.

%%%%%
\section{Analysis and Results} \label{sec:analysis}

The (Higgsino) kink track can be characterized by two types of features. First, a Higgsino kink track is usually isolated from other leptons or quarks. Second, the kink track from a new physics beyond the SM is likely a heavy particle while SM backgrounds will be from light particles. Meanwhile, the Wino DT is also characterized by similar features: no energetic particles along the DT direction and the heaviness of a DT. Thus, we consider the following observables, similarly to the latest LHC DT search observables: (1) $E_{\Delta R<0.5}$, total energy of all charged leptons and quarks within $\Delta R<0.5$ of the charged Higgsino, and (2) $p_T$(Higgsino). In addition, we look at (3) $\Delta R_{\rm min}$, the minimum $\Delta R$ between the charged Higgsino and any charged leptons or quarks, which can characterize charged activities along the original Higgsino direction.

We restrict to the events where the charged Higgsino decays within the relevant part of detectors: $30 {\rm cm} \leq L \lesssim 90 {\rm cm}$~\cite{Aad:2013yna,CMS:2014gxa}, where the $L$ is the decay length in the transverse plane. For this, each event generated by MadGraph5 v2.3.3 MSSM model~\cite{Alwall:2014hca} is reweighted by the probability for at least one Higgsino to decay within this part of detector. For the given transverse momentum $\gamma \beta$ and the lifetime $c \tau$ of the charged Higgsino, the probability for the Higgsino to decay within the part is
\beq
P = \exp \left(  -\frac{30{\rm cm} }{\gamma \beta c \tau} \right) \- \exp \left(  -\frac{90{\rm cm} }{\gamma \beta c \tau} \right).
\eeq
The probability for at least one Higgsino to decay properly can be obtained from this. For simplicity, we ignore longitudinal decay lengths which would not significantly change our results.

\begin{figure}[t] \centering
\includegraphics[width=0.45\textwidth]{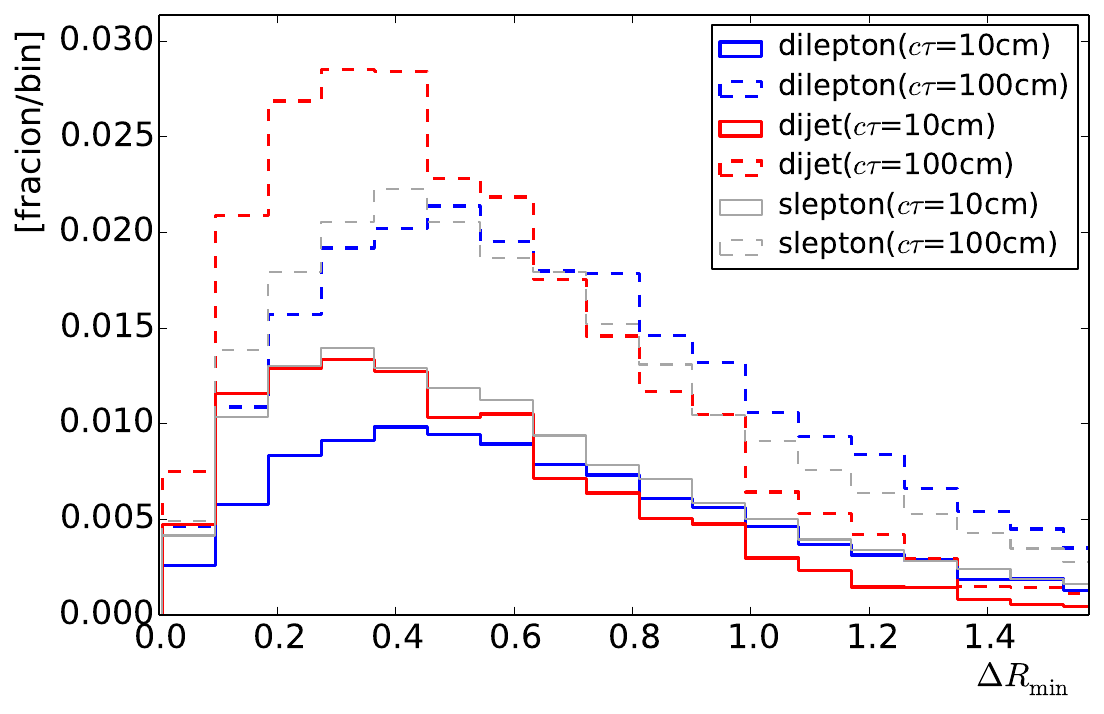}
\caption{$\Delta R_{\rm min}$, the minimum $\Delta R$ between the 300 GeV charged Higgsino and charged leptons and quarks. The dilepton channel (blue) where both $W$ bosons decay to leptons and the dijet channel (red) where both $W$ decay to quarks are compared. Higgsino lifetime $c\tau = $10 cm (solid), 100 cm (dashed) and the massless LSP. The fraction of events decaying within the appropriate part of detector is shown. CMS pre-selection cuts are applied. For comparison, slepton kinks (gray) are also shown.
}
\label{fig:dRmin}
\end{figure}

\begin{figure}[t] \centering
\includegraphics[width=0.45\textwidth]{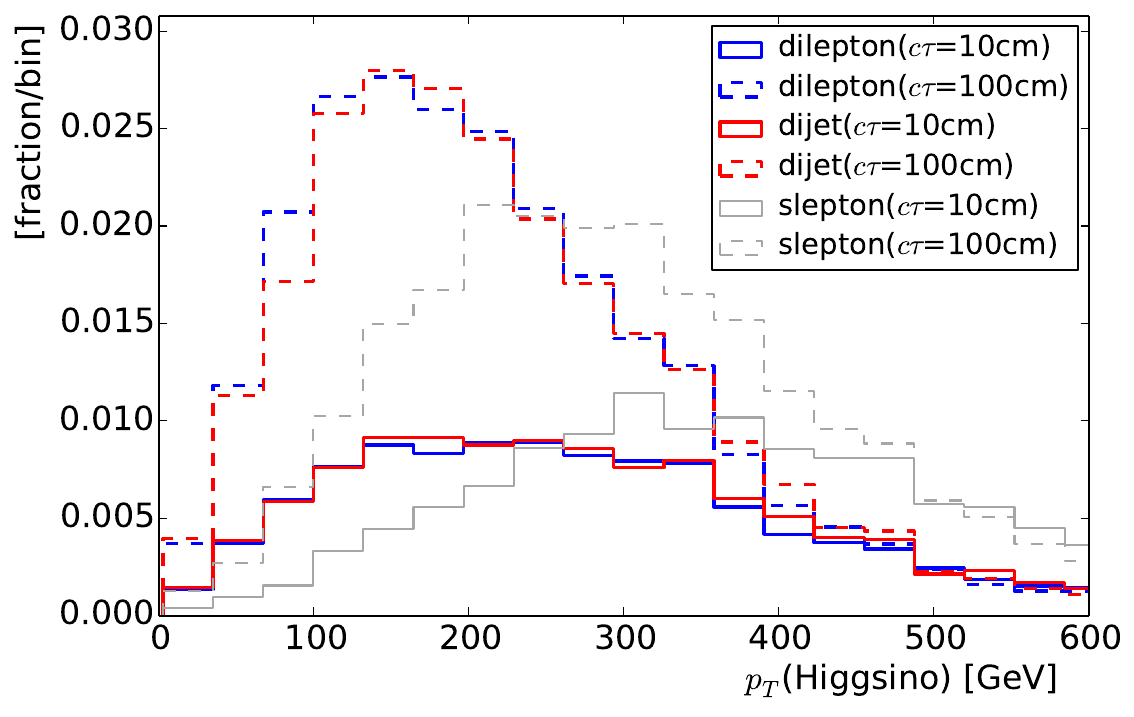}
\caption{$p_T$ of the charged Higgsino kink. Other details are as in \Fig{fig:dRmin}.
}
\label{fig:ptx}
\end{figure}

\begin{figure}[t] \centering
\includegraphics[width=0.45\textwidth]{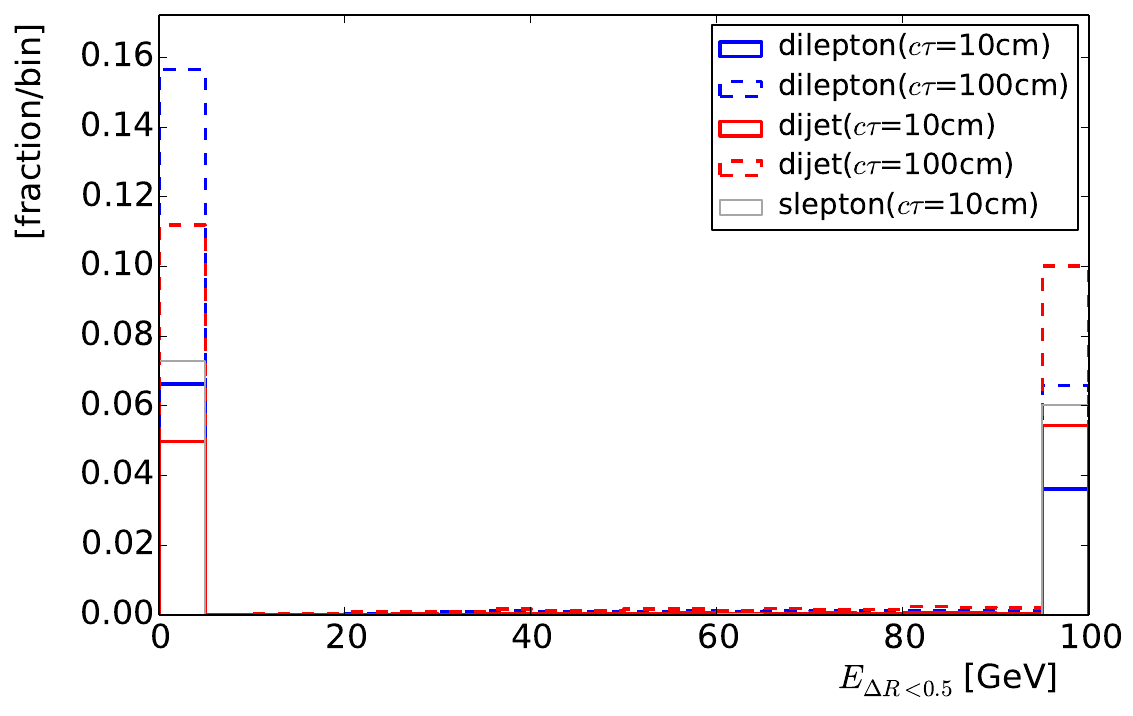}
\caption{$E_{\Delta R<0.5}$, total energy of all charged leptons and quark within $\Delta R<0.5$ of the charged Higgsino kink. Other details are as in \Fig{fig:dRmin}.
}
\label{fig:Ecalo}
\end{figure}

We show the resulting observables for the 300 GeV Higgsino and the massless LSP in \Fig{fig:dRmin}, \ref{fig:ptx}, and \ref{fig:Ecalo}. These are obtained after applying CMS pre-selection cuts (triggering the pair production of charginos plus initial-state radiation)~\cite{CMS:2014gxa}: at least one well-separated prompt jet and missing transverse energy (MET) with $p_T>110$ GeV, MET $>$ 110 GeV, $\Delta \phi$(jet, MET)$ > 0.5$, and no isolated prompt leptons. Although we base on the CMS analysis for our convenience, similar discussions can be made with the ATLAS analysis~\cite{Aad:2013yna}. Slepton kink events are also shown for comparison, where the long-lived slepton decays to the gravitino LSP and a charged lepton.

``$\Delta R_{\rm min}$'' shown in \Fig{fig:dRmin} measures charged activities along the original direction of the Higgsino kink. It is usually larger in the dilepton channel (where both $W$ bosons decay to leptons) than in the dijet channel (where both $W$ decay to quarks) because a smaller number of charged leptons and quarks are produced. 
Charged activities near the Wino DT is also small. Thus, the latest CMS DT search requires charged activities within $\Delta R<0.03$ of the DT to be very small: $\sum_{i \in \Delta R<0.03} p_T^i/ p_T({\rm DT}) <0.05$~\cite{CMS:2014gxa} (the scalar sum of all track $p_T$ compared to the $p_T$(DT)). To approximately mimic this cut, we require $\Delta R_{\rm min} >0.03$ in our final analysis; we can see that more than 95\% of Higgsino kink events can satisfy this cut.

$\Delta R_{\rm min}$ of the slepton kink may be expected to be similar to that of the dilepton channel since the same number of final-state charged leptons are produced. However, the slepton kink has somewhat smaller $\Delta R_{\rm min}$ as shown in \Fig{fig:dRmin} mainly due to the threshold suppression of the slepton scalar pair production as will be discussed.


Higgsino kinks have high $p_T \sim m({\rm Higgsino}) \sim {\cal O}(100)$ GeV as shown in \Fig{fig:ptx}. The $p_T$(Higgsino) distribution for the appropriate part of detector does not depend on the decay modes of the $W$ boson, but it depends sensitively on the chargino proper lifetime $c\tau$ as its decay length is (probabilistically) determined by $c\tau$ and its boost. The Higgsino kink and the Wino DT have the same $p_T$ distribution because both are Drell-Yan pair produced electroweak fermions; whereas, the slepton kink has harder $p_T$ distribution because a scalar pair Drell-Yan production suffers a threshold suppression due to angular momentum conservation. Thus, the $p_T$ distribution can be used to distinguish slepton kinks from others. The threshold suppression also explains why slepton kinks tend to have smaller $\Delta R_{\rm min}$ in \Fig{fig:dRmin}; harder sleptons with higher boosts lead to more collimated decay products. Following the latest CMS DT search, we apply $p_T$(Higgsino)$>50$ GeV in our final analysis; again, the majority of the Higgsino kink events satisfy this.

``$E_{\rm \Delta R<0.5}$'' shown in \Fig{fig:Ecalo} is the total energy of all charged leptons and quarks within $\Delta R=0.5$ of the Higgsino. Since the decay products of the chargino are energetic ($p_T \sim {\cal O}(100)$ GeV), either very small (if no charged decay products are nearby) or large (if at least one is nearby) value is predicted. The dilepton channel, having a smaller number of charged leptons, tends to have more events at small $E_{\rm \Delta R<0.5}$ than the dijet channel. On the other hand, the Wino DT always has small value of $E_{\rm \Delta R<0.5}$.  Following the latest CMS DT search, we require $E_{\rm \Delta R<0.5}<$ 10 GeV in our final analysis; still more than a half of Higgsino kink events would satisfy this.

\begin{figure}[t] \centering
\includegraphics[width=0.45\textwidth]{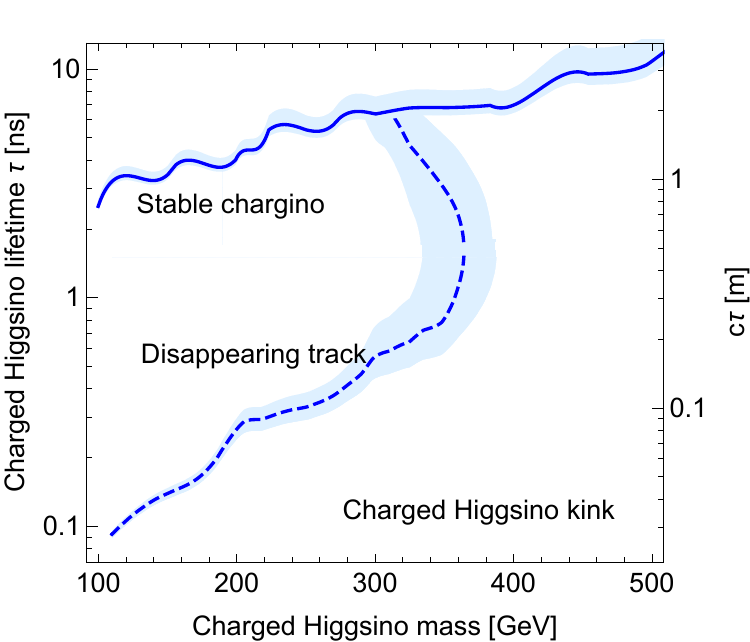}
\caption{Constraints on the Higgsino kink from 8 TeV DT~\cite{Aad:2013yna,CMS:2014gxa} and stable chargino searches~\cite{Khachatryan:2015lla} with 20/fb. The region left to each curve is constrained. The uncertainty band is obtained by varying selection efficiencies by 50\%.
}
\label{fig:constraint}
\end{figure}

From these considerations, we expect that a large fraction of Higgsino kink events would be captured by LHC Wino DT searches; in other words, our kink track that suddenly changes its direction (due to charged Higgsino's decays) can mimic DTs that actually disappear suddenly.
We derive constraints on the Higgsino kink by applying the final dicsovery cuts mentioned above. 
We first obtain the ratio of Higgsino kink efficiency to that of the Wino DT.\footnote{The number of background events reported in Ref.~\cite{CMS:2014gxa} is less than 10 for $m_{\widetilde{H}^+} \gtrsim 200$ GeV, so our approximation in this way is somewhat reliable.} For the given Higgsino mass, we then find the $c\tau$ of the Higgsino that can compensate the efficiency ratio and different effective chargino production rates. It is the constraint shown in \Fig{fig:constraint}.\footnote{We encourage a more dedicated study improving our parton-level results by including realistic detector effects, parton showering and reconstruction algorithms; our treatment of particle momentum is ideal and accurate efficiencies of our model can only be obtained by dedicated simulation studies.}

Constraints are strongest near $c\tau \simeq 50$ cm, excluding up to the 360 GeV Higgsino. There is no constraint for $c\tau \lesssim 3$ cm as the Higgsino has to decay after 30 cm; while a longer lifetime $c\tau \gtrsim 1$ m is already constrained by stable chargino searches~\cite{Khachatryan:2015lla}. The blue band is obtained by varying our efficiency by 50\%. We have assumed the massless LSP in this figure. But current CMS (and ATLAS too) selection efficiencies do not depend strongly on the mass difference between the NLSP and LSP as long as the $W$ boson can be on-shell; the massless LSP case has only 10\% larger efficiency than that of the compressed case. Thus, similar results will be obtained for other LSP masses.

In comparison with the Wino DT, the efficiency of the Higgsino kink is usually about the half of the Wino DT's with the same mass; somewhat smaller (larger) for the lighter (heavier) Higgsino. $E_{\Delta R<0.5}$ in \Fig{fig:Ecalo} is a main difference for this. The total effective charged Higgsino pair production rate (sum of all possible neutral and charged Higgsino pair productions in our model; see \Sec{sec:model}) is about 20\% smaller than the effective Wino DT production rate. In all, the maximum reach is reduced from about 520 GeV for the Wino DT~\cite{CMS:2014gxa} to 360 GeV for the Higssino kink. But the strongest constraints are similarly obtained for $c\tau \sim {\cal O}(10)$ cm due to similar $p_T$(Higgsino/Wino) distributions.

\begin{figure}[t] \centering
\includegraphics[width=0.45\textwidth]{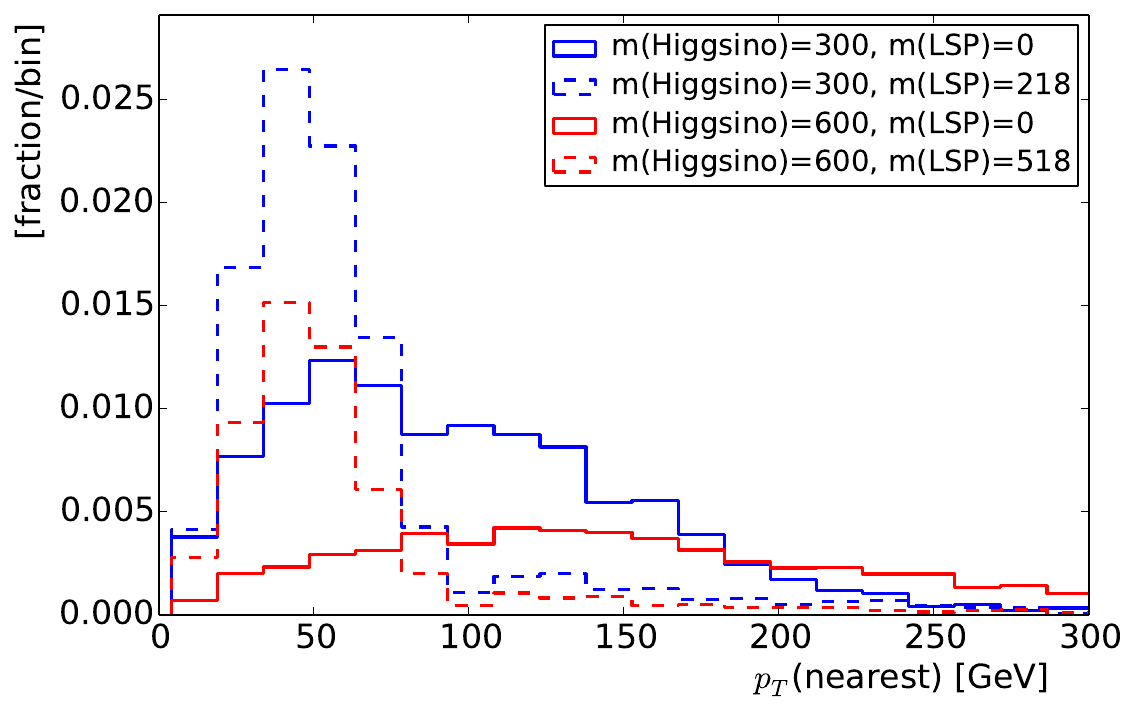}
\caption{$p_T$ of the charged lepton or the quark nearest to the charged Higgsino kink after all CMS discovery cuts. It is sensitive to the NLSP-LSP mass difference. $m$(LSP) $=0$ (solid) and compressed case with $m$(Higgsino)$- m$(LSP) $\approx m_W$ (dashed) are shown for $m$(Higgsino) $=$ 300 (blue) and 600 GeV (red). All $W$ decay modes are summed. 
}
\label{fig:pt-nearest}
\end{figure}

The NLSP-LSP mass difference of the Higgsino model can be apparent in the $p_T$ of the charged lepton or the quark nearest to the Higgsino. This reasonably measures the energy of chargino decay products, and it directly probes the mass difference. It is because the smaller the mass difference, the softer the visible decay products.
 This is clearly shown in \Fig{fig:pt-nearest}. This result is obtained after all CMS discovery cuts, so presumably backgrounds are already small there. Thus, this distribution can be useful in inferring mass parameters or in distinguishing Higgsino kink versus Wino DT that does not accompany energetic charged particles nearby.

%%%%%
\section{Conclusion} \label{sec:conclusion}

We have constrained the supersymmeric model of the long-lived charged Higgsino NLSP and the weakly interacting gravitino LSP from LHC searches. For $c\tau \sim 50$ cm, up to the 360 GeV Higgsino can be excluded by the latest 8 TeV LHC DT searches. But we encourage a more dedicated study by including various realistic effects that our parton-level results do not take into account. We have also discussed similarities and differences among the Higgsino kink versus the much-discussed Wino DT and the slepton kink. The differences are most apparent in observables measuring the activity nearby DT/kink (for Higgsino kink versus Wino DT) and in the $p_T$ of DT/kink (for Higgsino kink versus slepton kink). Moreover, we have discussed that the $p_T$ of the charged lepton or the quark nearest to the Higgsino can be used to infer the NLSP-LSP mass difference. Similar discussions can also be applied to various other models including the weakly interacting axino LSP~\cite{Martin:2000eq,Freitas:2011fx,Barenboim:2014kka} and $R$-parity violating slow decays (through $LLE^c$ and $LQD^c$ operators) of the charged LSP.

%%%%%%%%%%%%%%%
%\vspace{5mm} 
{\it Note added:} After the submission of the first draft to arXiv, more experimental information became available, and we have revised our study significantly. Some conclusions and focus are changed accordingly.

{\it Acknowledgements.}
We thank Brian Batell, Suyong Choi, Andy Haas, Teruki Kamon, and Jianming Qian for discussions.
SJ thanks the Aspen Center for Physics where part of the work was done.
HL appreciates the hospitality during his visit to KIAS.
The work of SJ was supported in part by National Research Foundation of Korea (Grant 2013R1A1A2058449) and US Department of Energy under contract DE-AC02-76SF00515.
The work of HL was supported in part by CERN-Korea fellowship and IBS (Project Code IBS-R018-D1).

%%%%%%%

\end{document}